\documentclass[12pt]{article}

\usepackage{amsmath}
\usepackage{graphicx,psfrag,epsf}
\usepackage{enumerate}
\usepackage{natbib}
\usepackage{url}

\newcommand{\blind}{1}

\begin{document}
	
	\def\spacingset#1{\renewcommand{\baselinestretch}%
		{#1}\small\normalsize} \spacingset{1}


\if1\blind
{
	\title{\bf Non-Parametric Cluster Significance Testing with Reference to a Unimodal Null Distribution}
	\author{Erika S. Helgeson \thanks{
			This material is based upon work supported by the National Science Foundation Graduate Research Fellowship under Grant No. DGE-1144081. 
			}\hspace{.2cm}\\
		Department of Biostatistics, University of North Carolina\\
		and \\
		Eric Bair \\
		Department of Biostatistics and Endodontics, University of North Carolina}
	\maketitle
} \fi

\if0\blind
{
	\bigskip
	\bigskip
	\bigskip
	\begin{center}
		{\LARGE\bf Title}
	\end{center}
	\medskip
} \fi

\bigskip
\begin{abstract}
Cluster analysis is an unsupervised learning strategy that can be employed to identify subgroups of observations in data sets of unknown structure. This strategy is particularly useful for analyzing high-dimensional data such as microarray gene expression data. Many clustering methods are available, but it is challenging to determine if the identified clusters represent distinct subgroups. We propose a novel strategy to investigate the significance of identified clusters by comparing the within-cluster sum of squares from the original data to that produced by clustering an appropriate unimodal null distribution. The null distribution we present for this problem uses kernel density estimation and thus does not require that the data follow any particular distribution. We find that our method can accurately test for the presence of clustering even when the number of features is high.
\end{abstract}

\noindent%
{\it Keywords:} Hierarchical Clustering; High-Dimension, Low-Sample Size Data; K-Means;  Statistical Significance;  Unimodality; Unsupervised Learning 
\vfill

\newpage
\spacingset{1.45} 
\section{Introduction}
\label{sec:intro}

In an initial analysis of a data set it is often of interest to discover if there are any natural groupings present in the data. Graphical visualizations may be useful, but for high dimensional data this is infeasible. Clustering is a common tool used for analyzing high dimensional, complex, data sets. This is an unsupervised method where patterns in the data are identified without specifying a response variable or building a model. Clustering is very versatile and can be applied to any data set in which similarities between observations can be measured. Several commonly used methods include hierarchical clustering, k-means clustering, and spectral and graph-based methods The exploratory nature of cluster methods have been proven to be very useful for identifying patterns in many fields and has been especially useful in the field of bioinformatics.

After clusters have been identified, the next logical step in the research process is to assess the validity of these putative clusters to determine if they truly represent distinct subgroups. Since many clustering algorithms will group the data into clusters even if the data is homogeneous, a cluster validation step is pivotal. 

Several methods have been developed which assess the strength of identified clusters. These methods generally test the null hypotheses that the data is homogeneous by comparing a statistic from the observed data to the statistic from an appropriate null distribution. The null distribution is generated under certain assumptions such as a specific parametric distribution or non-parametric assumptions about cluster shape. 

In the this paper we develop a novel non-parametric cluster significance test, UNPCI (unimodal non-parametric cluster index) test, with application to high-dimension-low-sample size (HDLSS) data. Our method is based on the definition of a cluster as data coming from a single unimodal distribution. We test the hypothesis that the data contains more than one cluster by comparing the cluster index, CI, from the data to the CI from an appropriate unimodal null distribution.  The permutation p-value can be used as an indication of the strength of the clustering in the data. Our method is versatile in a wide variety of settings including situations were normality assumptions do not hold.

The article is organized as follows. We first present the algorithm and details for the UNPCI method. Next,  we present theoretical properties of our proposed method. Then we briefly discuss alternative cluster significance testing methods and compare these methods to the UNPCI test in simulation studies We also present an application of our method to real data. We finish with a discussion of our method and simulation results. 

\section{Methods}
\subsection{The Development of the UNPCI test } \label{ourmethod}
To introduce the UNPCI test assume the data set can be expressed in the form of a $n \times p$ matrix $\emph{X}$, with $n$ observations and $p$ covariates or features. For ease of explanation assume the data is centered such that each feature has a mean of zero. As mentioned previously, we define a cluster as a single unimodal distribution where a mode is a single point of highest density. Our null and alternative hypotheses can be represented as:

\hspace{1.5cm}	$H_0$: The data come from a single unimodal distribution.

\hspace{1.5cm}  $H_a$: The data do not come from a single unimodal distribution.

When $H_0$ is not rejected we conclude that we have no evidence against the assumption that the data comes from a unimodal distribution. Thus we are unable to conclude that the given clustering of the data is real.

To perform this hypothesis test we need to approximate $\emph{X}$ under the null hypothesis of unimodality. The goal is to generate a null data set $\emph{X}^0$ as close to the original data set as possible under the restriction of unimodality. To achieve this goal without imposing additional parametric constraints a Gaussian kernel density estimator (KDE) is used for each feature. For a given feature, $j$, the KDE can be expressed as:
\begin{equation}
\hat{f}_j(t;h_j)=(nh_j)^{-1}\Sigma_{i=1}^nK(h^{-1}_j(t-X_{ij}))
\end{equation}
Where $h_j$ is the bandwidth; $K(\cdot)$ is the Gaussian kernel function; and $X_{1j},...,X_{nj}$ are the entries for feature j. \citet{Silverman1981using} showed that that a critical bandwidth, $h_{kj}$, can be determined such that 
\begin{equation}
h_{kj}=inf\{h_j:\hat{f}_j(\cdot;h_j) \ has \ at \ most \ k \ modes\}.
\end{equation}
We find $h_{1j}$ and rescale $\hat{f}_j(t;h_{1j})$ to have variance equal to the sample variance. Bootstrap samples from this rescaled KDE are generated by:
\begin{equation} \label{nulldist}
X_{ij}^0=(1+h_{1j}^2/\sigma_j^2)^{-1/2}(X_{I(ij)}+h_{1j}\epsilon_i)
\end{equation}
Where $\epsilon_i \sim N(0,1)$, $\sigma^2$ is the sample variance for feature j, and $X_{I(ij)}$ are sampled uniformly, with replacement, from the observed data for feature j. We use the notation N$(\mu, \sigma)$ to represent a normal distribution with mean $\mu$ and standard deviation $\sigma$. 

\begin{figure} 
	\centering
	\includegraphics[scale=0.5]{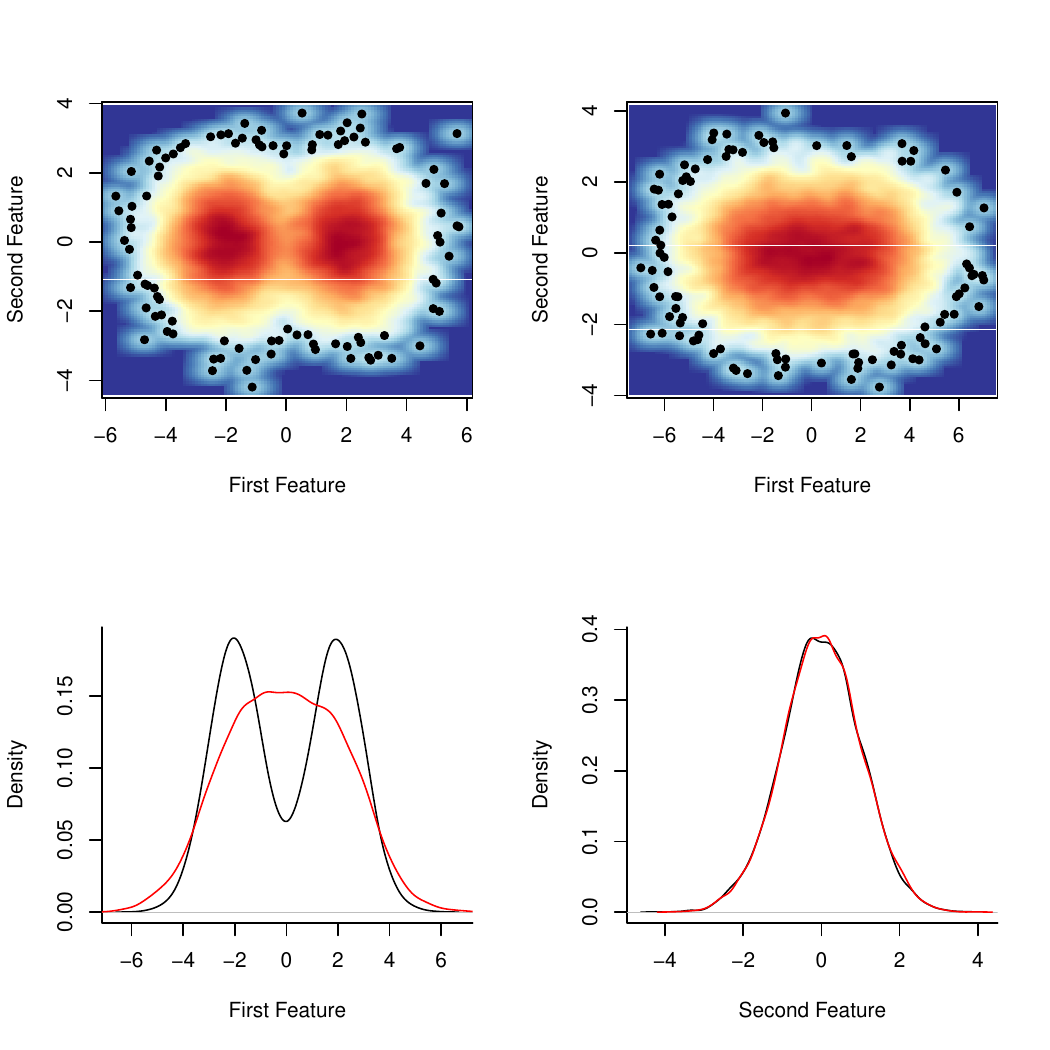}
	\caption{\small \textit{Illustration of the proposed reference distribution.} In this example two clusters are present in the data and the clusters differ only in the mean of the first feature. The first row gives the bivariate distribution for the observed data (left) and the reference distribution (right). The second row gives the univariate density for the first feature (left) and second feature (right). The black lines are from the observed data and the red lines are from the reference distribution. Notice the null distribution is bivariate unimodal and also unimodal in both features. It also very closely approximates the density of the second feature. } \label{F:Simple}
\end{figure}

Figure \ref{F:Simple} illustrates the effectiveness of our method in producing a unimodal distribution that closely approximates the observed data distribution. In the example given in this figure the two clusters differ in the mean of the first feature. Both clusters have the same distribution for the second feature. 

Even though the methodology applied will give a close unimodal approximation for each features, the multivariate data structure will be very different from the observed structure. If one were to naively set $X^0=\{X_1^0,...,X_p^0\}$ the original covariance structure of $X$ would not be preserved. To ensure that the first and second moments from the reference distribution approximate the moments from the observed data we propose that $\emph{X}$ be scaled such that the features have variance equal to one before generating $X^0$ using the methodology described above. Then  $X^0$ can be multiplied by the Cholesky root of the estimated covariance from $X$ and the covariance structure in $X^0$ will be the same to the covariance structure in $\emph{X}$. 

If the number of features, $p$, is less than the number of observations $n$, a sample covariance estimate can easily be calculated. For the situation when $p>n$ we use the graphical lasso by \citet{Witten2011glasso} to approximate the covariance structure since it generates a sparse covariance matrix and it can be computed quickly. Further research is needed to determine if other covariance estimators would be better suited for our cluster significance testing method. 

The graphical lasso problem maximizes the penalized log-likelihood:
\begin{equation*}
\textup{log det} \ \Theta-\textup{tr} (S\Theta)-\rho||\Theta||_1
\end{equation*}
over nonnegative definite matrices $\Theta$. Here tr denotes the trace, $||\Theta||_1$ is the sum of the absolute values of the elements of $\Theta$, the estimate of the inverse covariance matrix is given by $\Theta$, S is the empirical covariance matrix, and $\rho$ is a tuning parameter responsible for controlling the amount of $\ell_1$ shrinkage. The authors do not give specifications for choosing $\rho$, but we have found, in practice, that $\rho=0.02$ produces good results for most applications. 

The feature by feature Gaussian KDE methodology given above can become very time consuming as the number of features increases. In the high dimensional setting we propose using a dimension reduction technique similar to the method \citet{Bair2004Semisupervised} used to select a subset of the predictors based on their association with the outcome. We chose a subset of features such that their association with the binary cluster assignment  is strong. In this paper we chose features such that the p-value for the t-test for difference in means between the two groups is less than $\alpha$=0.10. Alternatively a set number of features can be chosen. 

To measure the strength of clusters in the observed data we use the two-means cluster index, CI, which is the ratio of the within cluster sum of squares  and the total sum of squares. 
\begin{equation}
CI= \frac{\Sigma^2_{k=1}\Sigma_{j\in C_k} ||x_j- \bar{x}^{(k)}||}{\Sigma^n_{j=1}||x_j-\bar{x}||}
\end{equation}
Here  $\bar{x}^{(k)}$ is the mean of cluster k for k=1 and 2, $C_k$ is the sample index for the kth cluster, and $\bar{x}$ is the overall mean. Smaller values of the CI indicate that a larger value of the overall variation is explained by the clustering.

We now summarize the procedure for the UNPCI test as follows:
Assume that $\emph{X}$ has been centered so the feature means are equal to zero.

\begin{enumerate}
	\item Identify putative clusters in data set. Unless otherwise noted, we apply k-means clustering with $\emph{k}$=2 after scaling the features to have variance equal to one. This scaled data set will be referred to as $\emph{X}^s$.  \label{clusterstep}
	\item Optional dimension reduction for high dimensional data sets. \label{dimeredstep} 
	\begin{enumerate}
		\item Identify features of $\emph{X}$ which are strongly associated with cluster identification. 
		\item Call the data set containing only the subset of features $\emph{X}^*$.
		\item Rerun cluster identification algorithm on data set $\emph{X}^*$. \label{dimclust}
	\end{enumerate}  
	\item Calculate the two-means clustering index for $X$ (or $X^*$), $CI_{data}$.
	\item Calculate an estimate for the covariance of $\emph{X}$ (or $X^*$). For the situation when $p>n$ We use the graphical lasso of \citet{Witten2011glasso} with sparsity parameter $\rho=0.02$.
	\item Generate multivariate unimodal reference distribution, $X^0$. The following will be repeated B times. We use B=1000 simulations in this paper.
	\begin{enumerate}
		\item For each feature in $\emph{X}^s$ (or $X^*$ scaled such that each feature has unit variance) find the smallest bandwidth estimator such that the Gaussian KDE of the feature has one mode. 
		\item For each feature generate data using a sample from the unimodal Gaussian KDE. 
		\item Multiply samples by Cholesky root of estimated covariance matrix to generate $X^0$.
		\item Cluster $X^0$ using the same clustering algorithm in step \ref{clusterstep} (or step \ref{dimclust})
		\item Calculate the cluster index, $CI_b$, for null data generated in iteration b.
	\end{enumerate}
	\item Calculate permutation p-value as follows: $\sum_{b=1}^B \{CI_b>CI_{data}\}/B$
	\item Alternatively a normalized p-value can be calculated by comparing the following z-score to the standard normal distribution. $Z= (CI_{data}-\mu_{CI})/\sigma_{CI}$ where $\mu_{CI}$ and $\sigma_{CI}$ represent the mean and standard deviation of the null CIs, respectively. 		
\end{enumerate}
We can conclude that a test is statistically significant if the p-value is less than a pre-specified level, $\alpha$, typically chosen to be 0.05. 
\subsection{Hierarchical Clustering} \label{hierarchical}
One strength of our proposed method is its application to a variety of clustering methods. All that is required to perform our method is a cluster identity for each observation and a measure of distance about the cluster means. Given this information, the proposed method using the cluster index is able to detect if the observations are clustered more closely together than what would be expected from a unimodal distribution. Thus the method can also be used to test the significance of clusters identified through a hierarchical clustering method. It should also be noted that the $L_2^2$ distance used in the cluster index could easily be replaced with the $L_1$ distance or other distance measure in order to accommodate different data structures and assumptions. 

To implement our proposed method for hierarchical cluster significance testing, simply use hierarchical clustering methods to generate cluster labels in step \ref{clusterstep}. For the hierarchical clustering examples in this paper the Euclidean distance matrix is calculated and single linkage or Ward's minimum variance, (Ward, 1963), hierarchical clustering methods were applied to the distance matrix. The produced tree was cut to give two clusters.

\subsection{Theoretical Properties}
 We now seek to establish several important theoretical properties for our reference distribution and cluster index test statistic. \citet{tibshirani2001estimating} show that for their gap statistic, a statistic also based on the sum of squares, that there is no least favorable multivariate unimodal reference distribution without degenerate support. Also they state that a maximum likelihood estimate of a multivariate unimodal reference distribution can not be shown to exist. Thus we we will not be able to choose a generally applicable reference distribution. However we can show that our reference distribution has optimal characteristics such as convergence in first and second moments and multivariate unimodality. We also seek to show asymptotic convergence of the CI from the reference distribution to the data CI when the data in unclustered and asymptotic divergence when the data is clustered.

  When $n>p$ it is trivial to show that the first and second moments of our proposed reference distribution asymptotically approach the first and second moments of the observed data. When $p>n$ showing the convergence of second moment depends on the method used to estimate the covariance (in this paper the graphical lasso is used)  and is beyond the scope of this paper. 
   
   From \citet{Silverman1981using} we have that $ h_{1j}$ can be chosen such that $\hat{f}(\dot,h_{1j})$ has at most one mode. Thus before multiplying by the Cholesky root each feature in our reference distribution is unimodal. To show that the marginal distributions retain their unimodality and the joint reference distribution is multivariate unimodal we appeal to the concept of \textit{strong unimodality}. \citet{Ibragimov1956on} defines a distribution function to be $\emph{strong unimodal}$ if its composition with any unimodal distribution function is unimodal. We show that the features are strong unimodal and thus the final marginal distributions are unimodal. 
   \\
   
      \textit{Theorem 1}. Each feature in the reference distribution is strong unimodal. 
      \\
      
  We now need to show that our transformed distribution is multivariate unimodal. To do so we will use the definition 2.5 from \citet{Sager1978Estimation}. Let $\boldsymbol{x}$ be a point in p-space $\boldsymbol{x}=(x_1,..,x_p)$ and $d(\boldsymbol{x},\boldsymbol{y})$ be the Euclidean metric $|\boldsymbol{x}-\boldsymbol{y}|$. He defines a point $\boldsymbol{\theta}$ to be a  multivariate mode of $F$ if for each $\epsilon>0$, $\exists \ \delta >0$ s.t. $d(\boldsymbol{x},\boldsymbol{\theta})> \epsilon$ implies $f(\boldsymbol{x})$+ $\delta < f(\boldsymbol{\theta})$.
  \\
  
  \textit{Theorem 2}. Suppose that each feature $x_j$ is independent and unimodal with mode $max(f_j(x_j))=m_j$ and the unique mode for $f(x_1,..,x_p)$ is given by:  
  \begin{math}
  \underset{\boldsymbol{x}}{max} \{\ f(x_1,..,x_p) \} =<m_1,...,m_p>
  \end{math}. Then after multiplying by the Cholesky root of the covariance matrix the resulting distribution, $h(y_1,...,y_p)$, is multivariate unimodal.  
    \\
  
  We now show asymptotic convergence of our null CI to the data CI when the data is unclustered and divergence when the data is clustered. As $n \rightarrow \infty$ the sample data becomes continuous so our proof compares the theoretical cluster index, the cluster index assuming continuity in the data, from the null distribution to the theoretical distribution from the data. The theoretical cluster index is defined to be the theoretical within cluster sum of squares divided by the theoretical total sum of squares.  The within cluster sum of squares is based off of a partition of the feature space into two non-overlapping subspaces $S_1$ and $S_2$. Using the same notation as \citet{Huang2015statistical} the theoretical sum of squares is given by: 
   \begin{equation}
   TCI=\frac{WSS}{TSS}=\frac{WSS_1+WSS_2}{TSS}
   =\frac{\int_{\boldsymbol{x} \in S_1} ||\boldsymbol{x}-\boldsymbol{\mu_1}||^2\phi(x)dx +\int_{\boldsymbol{x}\in S_2} ||\boldsymbol{x}-\boldsymbol{\mu_2}||^2\phi(x) dx}{\int||\boldsymbol{x}||^2\phi(x)dx}
      \end{equation}
      Where $\boldsymbol{\mu_k}=\int_{\boldsymbol{x}\in S_k} \boldsymbol{x}\phi(x)dx$
      \\
   
   \textit{Theorem 3}.   Let $\boldsymbol{x}= (x_1,\dots, x_p)$ be a d-dimensional random vector having a multivariate normal    distribution of $ \boldsymbol{x} \sim N(\boldsymbol{0},\boldsymbol{D})$ where $\boldsymbol{D}$ is a known covariance matrix with diagonal entries $\lambda_1 \ge \lambda_2 \ge \dots \ge \lambda_p$. Let $TCI_{GAUSS}$ represent the theoretical cluster index of $\boldsymbol{x}$.  For the choice of $S_1$ and $S_2$ which minimizes WSS, \begin{math}
    TCI_{GAUSS}=1-\frac{2}{\pi}\frac{\lambda_1}{\sum_{j=1}^p\lambda_j}
    \end{math}. The theoretical cluster index for our null distribution, $TCI_{null}$ approaches $TCI_{GAUSS}$ as n $ \rightarrow \infty$.
    \\
    
     \textit{Theorem 4}.
   Let $x = (x_1,\dots, x_p)$ be a p-dimensional random vector having a mixture multivariate normal distribution. Specifically the data is distributed as $\eta f(x)+(1-\eta)g(x)$ where $f(x) \sim N(\boldsymbol{0},\boldsymbol{D})$, $g(x) \sim N(\boldsymbol{\mu},\boldsymbol{D})$,  $\eta \in (0,1)$ is the mixing proportion, $\boldsymbol{\mu}=(a,\dots,a)^T$ with nonzero constant a, and $\boldsymbol{D}$ is a diagonal matrix with elements $\lambda_1 \ge \lambda_2 \ge \dots \ge \lambda_p$. Suppose  the mixing proportion $\eta$ stays constant as $n \rightarrow \infty$. Also assume that the clustering algorithm is able to correctly classify the $n_1$ observations that arise from the $N(\boldsymbol{0},\boldsymbol{D})$ distribution into cluster 1 and the $n_2$ observations that arise from the $N(\boldsymbol{\mu},\boldsymbol{D})$ distribution into cluster 2. Let $TCI_{mix}$ represent the theoretical cluster index from the mixture distribution. Then we have that $\underset{n \rightarrow \infty }{\lim}TCI_{mix}<\underset{n \rightarrow \infty }{\lim}TCI_{null}$.
   \\
   
   Up to this point we have focused on $n \rightarrow \infty$ asymptotics for the cluster index. We now establish asymptotic properties as $p \rightarrow \infty$.
   \\
   
   \textit{Theorem 5}.
    Let $X$ be a mixture of two Gaussian distributions $X=\eta N(\boldsymbol{0},\boldsymbol{D})+ (1-\eta)N(\boldsymbol{\mu},\boldsymbol{D})$ where $X$ is an $ n \times p$ matrix and $\eta \in (0,1)$ is the mixing proportion. Let $n_1$ be the number of observations collected from $N(\boldsymbol{0},\boldsymbol{D})$ and $n_2$ be the number of observations collected from $N(\boldsymbol{\mu},\boldsymbol{D})$, Define $\boldsymbol{\mu}=(a,...,a)^T$ with a $\ne$ 0. Suppose $\boldsymbol{D}$ is a known diagonal covariance matrix with elements $\lambda_1>...>\lambda_p$. Note that the variance for feature j from the mixture distribution is given by $\lambda_j+\eta(1-\eta)a^2$. Assume n is fixed;  $min(n_1,n_2)>0$; $n_1+n_2=n \ge 3$; $\Sigma_{j=1}^p\lambda_j=O(p^\beta)$ with $0 \le \beta < 1$; $max_j(\lambda_j+\eta(1-\eta)\alpha^2) \le M$ with $M>0$ a fixed constant. Let the observed data with scaled and centered features be denoted as $X^s$. Also assume that a finite bandwidth $h_{1j}$, $max_j(h_{1j})<L$ with $L>0$ a fixed constant, can be chosen for each feature of $X^s$ such that the Gaussian kernel density estimator is unimodal. Then we can show that the corresponding p-value converges to 0 in probability as $ p \rightarrow \infty$.

\subsection{Existing Cluster Significance Testing Methods} \label{existingmethods}
Several cluster assessment strategies have been proposed with different clustering assumptions and measures of cluster strength.  The SigClust method of \citet{liu2008statistical} takes a parametric approach and defines a cluster as a single Gaussian distribution. To  tackle the problem of HDLSS covariance estimation they use a combination of invariance principles and a factor analysis model. They then compare the cluster index from the observed data to what would be expected from a single Gaussian distribution in order to test for clustering. The significance test of \citet{Maitra2012bootstrapping} avoids defining a parametric distribution, but assumes that clusters are compact and can be transformed to be spherical and similar to each other. Their bootstrapping method requires the estimation of the data covariance matrix and is not optimized for the HDLSS setting. We will refer to this method as BFS, bootstrapping for significance. \citet{Ahmed2012investigating} employ a different tactic altogether by first reducing the data set to be unidimensional by projecting the data onto its principle curve and then employing Silverman's bandwidth test (1981) to test for more than one mode. We will refer to this method as MPC, multimodality of principal curves. Instead of reducing the data to a univariate representation and performing a global modality test, \citet{Cheng2014MultivariateModality} use local tests to asses the significance of identified modes in a non-parametric estimate of the multivariate density.  The method of \citet{Kapp2006are} uses a validation measure called the in group proportion (IGP) based on the proportion of observations in a cluster whose nearest neighbors are also in the same cluster. They use the IGP to assess cluster reproducibility, the ability to find the same cluster in a different data set from which it was first identified, by comparing the IGP from the observed data to the IGP generated from an a null distribution in which observations are permuted within the box aligned with their principal components.

Although these methods have proven to be useful in certain settings each method has potential shortcomings. The SigClust method works very well when data are normally distributed. Since, by definition, it tests whether the data come from single multivariate Gaussian distribution it is not suited for cluster significance testing in non-normal settings. As pointed out above the cluster significance testing method of \citet{Maitra2012bootstrapping} does not rely on distributional assumptions, but due to difficulties in covariance estimation it cannot be used when the number of features exceeds the number of observations. Important data structures may be lost when the data is reduced to a unidimensional principal curve as in the method of \citet{Ahmed2012investigating}. Although the method of \citet{Cheng2014MultivariateModality} is a promising solution to the problem of multidimensional cluster significance testing, in practice we did it not find it comparable to our proposed method. Specifically, we were not able to test the significance of pre-identified clusters and the clusters the method found often only consisted of one observation. We will not be examining the results from this method in the following sections. Since the IGP method involves the permutation of all features within each observation it can be computationally expensive for large data sets and does not take into consideration covariance between features. Thus we have illustrated the need for a non-parametric cluster significance testing method that is optimized for the high dimensional setting. 

In the later sections we will compare our proposed cluster significance testing method to the methods described above. All k-means based methods were implemented in R version 3.2.2. R version 3.3.1 was used for the hierarchical clustering examples. Unless otherwise noted, k means with k=2 was first implemented on the scaled and centered data and then the same cluster identities were fed into each testing method. If k-means clustered the data such that only one observation belonged to a given cluster, the clustering was rejected and an additional simulation was run.  The proposed method utilizes the ``sparcl" v 1.0.3 and ``glasso" v 1.8 R packages. The SigClust v. 1.1.0 algorithm was used with 1000 simulations, and the covariance was estimated via soft-thresholding, sample covariance estimation, and/or hard-thresholding if appropriate. See \citet{Huang2015statistical}  for further discussion of covariance estimation for the SigClust method. When the number of features was less than the number of observations the bootstrapping method of \citet{Maitra2012bootstrapping} was used with 1000 replicates and it was assumed that the clusters were not homogeneous. The method of \citet{Ahmed2012investigating} was implemented by first using the R package ``princurve" v. 1.1-12 with the maximum number of iterations set to 100 and default parameter specifications. If convergence for the principal curve was not met the results were rejected. Then Silverman's bandwidth test (1981) with 10,000 bootstrap samples (per the author's recommendation) was implemented using code located at http://www-bcf.usc.edu/~gourab/code-bmt/tables/table-2/silverman.test.R. The IGP method of \citet{Kapp2006are} was implemented by first calculating centroids in the observed data set using the given cluster indices. These calculated centroids along with an additional data set with the same distribution as the observed data were then used in the IGP method implemented in the clustRepo v 0.5-1.1 package with 1000 permutations. Since the IGP method calculates reproducibility for each cluster, for comparative purposes, we calculated the number of times the p-values were less than 0.05 for both clusters. 

\section{Simulation Studies}
We evaluated the performance of our cluster significance testing method on a variety of simulated data set in comparison to the existing clustering methods discussed in section \ref{existingmethods}. Three data settings were considered: data sets with more observations than features (referred to as low dimensional); ``high dimensional"  data sets with 100 observations and 10,000 features, and two hierarchical clustering examples. For the low dimensional data sets we used the sample covariance in our generation of the null reference data. The graphical lasso (Witten et al., 2011) with $\rho$=.02 was used in the remaining two settings. The dimension reduction technique discussed in section \ref{ourmethod}  step \ref{dimeredstep} was incorporated in our method for the high dimensional data sets. For low dimensional data sets 100 replications of each simulation were conducted. Due to computation time only 10 replications were run for each high dimensional data sets and only 50 replications of each hierarchical clustering example were conducted. The results are presented in section \ref{simresults}.

\subsection{Low Dimensional Data Set Simulations}
We first tested the accuracy of our method with four un-clustered examples. For the ``5d sphere" example features for 1000 observations were generated from a standard normal distribution on the surface of a five dimensional sphere. The``Null normal" example was constructed as a 200 observation by 100 feature data set with i.i.d. standard normal features. The  ``Null correlated" example was constructed as a 200 by 100 standard Gaussian matrix with the first 40 features having a covariance of 0.20. The ``Null t" example was constructed as a 200 by 100 data set with the features having i.i.d. $t_2$ distributions (this notation represents the t distribution with 2 degrees of freedom).

To test the power of our method in detecting when clusters were truly present we generated four data sets each containing two clusters. The first three simulations were constructed as 200 observation by 100 feature data sets. In the ``Normal clustered" simulation the background followed a standard normal distribution, but the first 30 features for 50 of the observations followed a N(2, 1) distribution (this notation represents the normal distribution with mean 2 and standard deviation 1). In the``T clustered" simulation the background followed a $t_2$ distribution, but the distribution was shifted by a non-centrality parameter of 12 for the first 30 features in 40 of the observations. To test the ability of our method to detect clusters when correlation was present a simulation was generated where the background was a standard Gaussian matrix except the first 40 features had a shared covariance of 0.20. An additional N(2,1) layer was added to features 45:74 for observations 1:50. This simulation will be referred to as ``Correlated clusters". 

The final clustered simulation, ``Elongated clusters", was generated as a 202 observation by 3 features data set. Observations 1:101 belonged to the first cluster and the background was generated by setting the first three features equal to $t$ where $t$ took 101 equally spaced values from -.5 to .5. A N(0,0.10) distributed noise term was then added to each entry.  Observations 102:202 were distributed similarly except the constant value of 4 was added to each entry. 

\subsection{High Dimensional Data Set Simulations}
We tested the accuracy of our method in high dimensions for three null simulations. The ``Null normal" example was simply an independent standard Gaussian matrix. In the ``Null correlated" simulation the features were normally distributed with a mean of 0 and an AR(1) banded covariance matrix such that each feature had a variance of 0.80 and the covariance between feature $i$ and feature $j$ was defined as $cov(i,j)= I(|j-i|<42)(.80)^{|j-i|}$. In the ``Null t" example the features were i.i.d. $t_2$ distributed.

Two clustered simulation examples were generated for the high dimensional setting. The ``Normal clustered" example had i.i.d. standard normal distributed features except for 30 observations whose first 50 features had a mean of 2. The ``T clustered" example had i.i.d. $t_2$ distributed features except for 30 observations whose first 100 features had a non-centrality parameter of 12. 

\subsection{Hierarchical Clustering Data Set Simulations}
To illustrate the usefulness of our proposed method in testing the significance of clusters identified through hierarchical clustering methods we generated null and clustered data sets and then identified clusters using Ward's minimum variance method (Ward, 1963) and single linkage, respectively. Ward's method was used for the null data because single linkage did not identify any spurious clusters. The identified clusters were then tested using the procedure outlined in section  \ref{hierarchical}. One strength of our method is that it is not limited to only assessing clusters identified through k-means based approaches, but is flexible enough to handle other clustering methods.

The null example was a 500 observation by 75 feature data set generated as follows:
\begin{align*}
\ if \ 1 \le j \le 25 \left\{
\begin{array}{ll}
X_{i,2j}=-2 +5*U_{i,2j}
\\
X_{i,2j-1}=5 +5*U_{i,2j-1}
\end{array} \right.
\\ otherwise \ X_{ij} \sim N(0,1)
\end{align*}
Where the $U_{ik}$'s are i.i.d. Uniform(0,1) for k=1,...,50.

The clustered example contained 1200 observation with 75 features and was simulated as follows:
\begin{align*}
\ if \ 1 \le j \le 25 \left\{
\begin{array}{ll}
X_{i,2j}=-2*I(i \le 500) +5*sin(\theta_i +\pi I(i>500)) + \epsilon_i\\
X_{i,2j-1}=5*I(i \le 500) +5*cos(\theta_i +\pi I(i>500)) + \epsilon_i
\end{array} \right.
\\ otherwise \ X_{ij} \sim N(0,1)
\end{align*}

\begin{figure} 
	\centering
	\includegraphics[scale=0.35]{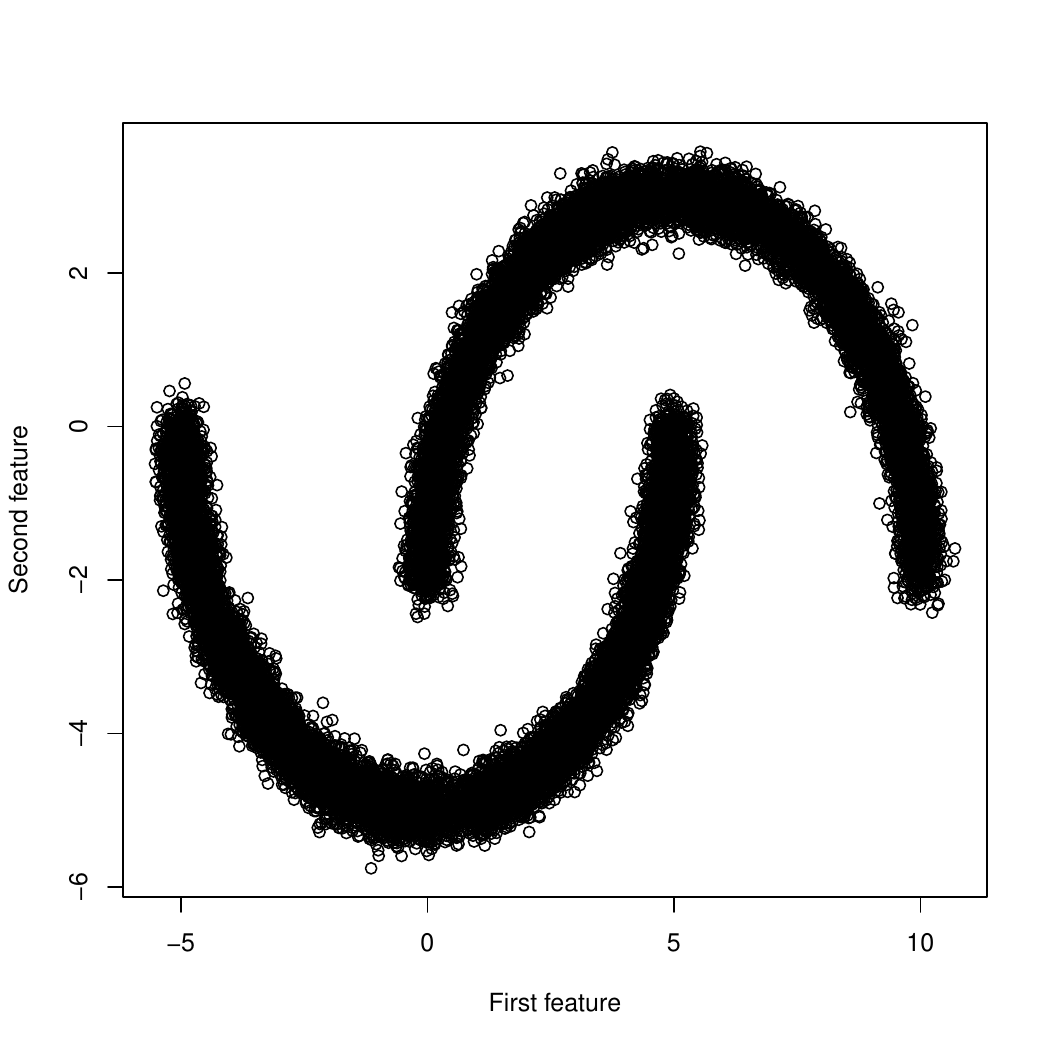}
	\caption{\small \textit{Hierarchical simulation example}. Plot of the second feature versus the first feature for a single simulation from the Hierarchical simulation scenario. Note that the data forms two non-spherical clusters. The UNPCI method detected that clusters were present in all of the 50 simulations.} \label{F:hierarchical}
\end{figure}

Where the $\epsilon_i$'s are i.i.d. N(0,0.2) and the $\theta_i$'s are i.i.d. Uniform(0,$\pi$).  Figure \ref{F:hierarchical} gives a visual representation of the cluster structure for one iteration of the simulation. 

An extension of SigClust specifically applied to the hierarchical setting has been proposed (https://arxiv.org/pdf/1411.5259.pdf). Since this method has not yet been published we compare the results of this method (implemented from R code located at 
https://github.com/ pkimes/sigclust2) to the published SigClust method which uses k-means clustering. Both data sets were simulated 50 times and the results are given in table \ref{T:Hresults}.

\begin{table}[ht] 
	\footnotesize
	\caption{\small Comparison of prediction accuracy for low dimensional clustering examples. SigClust 1, SigClust 2, and SigClust 3 represent SigClust implemented using the following covariance estimation methods: soft-thresholding, sample covariance estimation, and hard-thresholding, respectively.  The number of times each method gave a p-value $<0.05$ for 100 simulations is recorded.}
	\centering
	\begin{tabular} {l  c c c c c c c c}
		\hline \hline
		Simulation Name  & \multicolumn{7}{c} {Number of Simulations with p-value $<0.05$}  \\
		&    UNPCI & SigClust 1 & SigClust 2 & SigClust 3  & IGP & BFS & MPC \\
		5 d sphere & 	7 &	NA	& 41 &	83 &	1 &	0 &	98	\\							
		Null normal & 	0 &	0 &	0 &	0 &	0 &	0 &	71 \\								
		Null correlated & 	0 &	30 &	0 &	89 &	98 &	0 &	27 \\							
		Null t & 	1 &	0 &	0 &	0 &	1 &	98 &	84		\\						
		Normal clustered & 	100 &	100 &	100 &	100 &	14 &	100 &	99 \\							
		T clustered  &	97 &	68 &	69 &	69 &	9 &	100 &	83	\\					
		Correlated clusters & 98 &	98 &	98 &	98 &	65 &	75 &	26 \\
		Elongated clusters &  100 & 	100	& 100 &	100 &	0 &	100 &	100 \\						
		\hline
		\hline
	\end{tabular}
	\label{T:ldresults}
\end{table}

\subsection{Simulation Results} \label{simresults}
The results of our low dimensional, high dimensional, and hierarchical simulations are given in tables \ref{T:ldresults}, \ref{T:hdresults}, and \ref{T:Hresults}, respectively. For the low dimensional examples our method had a very low detection of ``significant" clusters (defined as having a p-value less than 0.05) when no clusters were present. Only the BFS and IGP methods outperformed the UNPCI test in the 5 dimensional sphere example, but UNPCI result's were closer to the five significant results we would expect by chance using an $\alpha$=0.05. Since SigClust is defined to detect deviations from normality it was not a useful test for the five dimensional sphere example. All methods except MPC performed with perfect accuracy in the ``Null normal" example. In the ``Null correlated" example the UNPCI test, SigClust with sample covariance estimation, and BFS had perfect performance. SigClust with hard thresholding and the IGP method detected clusters in a very large number of the simulations. For the ``Null t" example the SigClust methods had perfect results in detecting that none of the simulations had clusters, the UNPCI test and IGP detected clusters in one out of 100 simulations, and BFS and MPC had a very large number of inaccurate results.  MPC tended to always detect clusters and lacked the specificity needed to identify when clusters were not present. 

The UNPCI test performed well in the low dimensional clustered examples as well. For the ``Normal clustered" example all methods except IGP performed with high accuracy BFS and UNPCI were the top performers in the ``T clustered" example followed by MPC and the SigClust methods. IGP performed very poorly in this example. The UNPCI test and SigClust were the strongest performers in the ``Correlated clusters" example giving significant results for nearly all simulations. MPC performed very poorly in this example. All methods except the IGP gave perfect results for the ``Elongated clusters" example.

UNPCI test's utility is especially noteworthy in the high dimensional examples. The UNPCI test along with SigClust had perfect accuracy in the ``Null normal", ``Null correlated", and ``Null t" examples. Again, MPC tended to produce low p-values in most of the simulations regardless of whether clustering was actually present. IGP had perfect performance in the null normal and null correlated examples, but had a slightly inflated cluster detection probability in the ``Null t" example. The UNPCI test and MPC were the obvious winners in the ``Normal clustered" and ``T clustered" examples. 

The UNPCI test had perfect performance in the hierarchical clustering example when clustering was present. The other clustering methods, besides SigClust with sample covariance estimation, had similar performance.  When the data came from a null distribution  the UNPCI test outperformed the other competing methods, but had an inflated type 1 error.

\begin{table}[ht] 
	\footnotesize
	\caption{\small Comparison of prediction accuracy for high dimensional clustering examples. SigClust 1, SigClust 2, and SigClust 3 represent SigClust implemented using the following covariance estimation methods: soft-thresholding, sample covariance estimation, and hard-thresholding, respectively. The number of times each method gave a p-value $<0.05$ for 10 simulations is recorded. The average number of features selected using the dimension reduction technique is also noted.} 
	\centering
	\begin{tabular} {l  c c c c c c c c }
		\hline \hline
		Simulation Name & Ave. number of   & \multicolumn{6}{c} {Number of Simulations with p-value $<0.05$}  \\
		& features selected  &   UNPCI & SigClust 1 & SigClust 2 & SigClust 3  & IGP  & MPC \\
		Null normal     &	1240.5 &	0  &	0 &	0 &	0 &	0 &	8 \\								
		Null correlated &	1524.3 &	0  &	0 &	0 &	0 &	0 &	8 \\								
		Null t	        &   1070.5 &	0  &	0 &	0 &	0 &	2 &	9 \\								
		Normal clustered&	1088.8 &	8  &	0 &	0 &	0 &	1 &	8 \\								
		T clustered     &	1003.0 &	10 &	3 &	1 &	3 &	4 &	10\\								
		\hline
		\hline
	\end{tabular}
	\label{T:hdresults}
\end{table}

\begin{table}[ht] 
	\footnotesize
	\caption{\small Comparison of prediction accuracy for hierarchical clustering examples. SigClust 1, SigClust 2, and SigClust 3 represent SigClust implemented using the following covariance estimation methods: soft-thresholding, sample covariance estimation, and hard-thresholding, respectively. HSigClust represents the extension of SigClust specifically applied to the hierarchical setting. The number of times each method gave a p-value $<0.05$ for 50 simulations is recorded} 
	\centering
	\begin{tabular} {l  c c c c c c c c c}
		\hline \hline
		Simulation Name  & \multicolumn{8}{c} {Number of Simulations with p-value $<0.05$}  \\
		&    UNPCI & SigClust 1 & SigClust 2 & SigClust 3 & HSigClust  & IGP & BFS & MPC \\ 
		Null & 12 & 50 &15 &50 &46 &46 &46 &50 \\								
		Two clusters &50& 50& 0 & 50 &49 & 50& 50& 50 \\										
		\hline
		\hline
	\end{tabular}
	\label{T:Hresults}
\end{table}  
\section{Analysis of OPPERA data}
We use data collected from the Orofacial Pain: Prospective Evaluation and Risk Assessment (OPPERA) study to illustrate the usefulness of our proposed method. This study has been described previously (\citet{Slade2011Study}). In brief, OPPERA is a prospective cohort study aimed at discovering causes for Temporomandibular disorder (TMD). In order to achieve this goal individuals with and without TMD were recruited to 4 US study sites and their pain sensitivity, psychological distress, and autonomic function were examined through a battery of self-completed questionnaires and clinical assessments. Individuals who were identified as TMD-free at the start of the study were followed for up to 5 years to determine if they developed incident TMD.

TMD is diagnosed based on specific painful conditions in the masticatory muscles and temporomandibular joint (\cite{Schiffman2014Diagnostic}), but several different etiological mechanisms could be responsible for this disorder. \citet{Bair2016ClusterPaper} identified three clinically important subgroups within the OPPERA study. The data which they analyzed consisted of 115 features scaled to have mean 0 and variance 1 collected from 1031 baseline TMD cases and 3247 controls. As a first step, they used the supervised cluster analysis approach of \citet{Bair2004Semisupervised} to select the 25 features which are most strongly associated with chronic TMD. Next, they used the gap statistic of \citet{tibshirani2001estimating}  on this reduced data set to determine that three subgroups were present. These subgroups are denoted as the Adaptive Cluster (1426 total individuals), the Global Symptoms Cluster (790 total individuals) and the Pain-Sensitive Cluster (2062 total individuals) based on their risk factor characteristics. 

We examined the strength of the putative clusters using our proposed methodology. We performed pairwise comparisons between each of the three clusters using the scaled data for the 25 features most strongly associated with TMD. For comparative purposes instead of presenting the permutation p-value for our tests we use the normal approximation to calculate p-values.
\begin{figure} 
	\centering
	\includegraphics{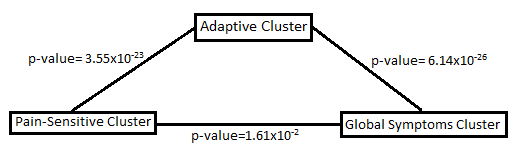}
	\caption{\small \textit{OPPERA cluster significance tests} Normal approximation p-values for testing significance of clusters identified in the OPPERA study} \label{F:OPPERA}
\end{figure} 

The results are given in figure \ref{F:OPPERA}. We find that all three clusters are well separated, but the Pain-Sensitive Cluster and the Global Symptoms Cluster are the most similar. This corroborates the results given by \citet{Bair2016ClusterPaper} who find that the three clusters are distinct with respect to their risk factor profiles. Specifically, the Adaptive Cluster has lower pain sensitivity and low psychological distress; the Pain-Sensitive Cluster has  high pain sensitivity, but low psychological distress; and the Global Symptoms Cluster with has high pain sensitivity and high psychological distress. Interestingly, the pain sensitive and global symptoms had higher proportions of TMD cases (26.2 percent and 51 percent, respectively) compared to 6.2 percent in Adaptive Cluster. Also individuals in the pain sensitive and Global Symptoms Cluster were more likely  to be female (72.9 percent and 69.6 percent) compared to 41.2 percent in the Adaptive Cluster. Even though individuals in the Global Symptoms Cluster were more likely to develop TMD than individuals in the other two clusters, the three clusters may show differing pain profiles once they develop TMD. Specifically, individuals in the pain-sensitive and Global Symptoms Cluster may be experiencing central sensitization whereas individuals in Adaptive Cluster may have more localized pain.  These findings illustrate the usefulness of our proposed significance test at assessing differences between clusters, an important preliminary step in identifying clinically meaningful subgroups.  

\section{Discussion}
Clustering is a machine learning tool that can be especially useful in discovering underlying structure in HDLSS data. Once putative clusters have been identified an important next step is determining if those clusters represent distinct subgroups. In this paper we have developed a non-parametric approach to test the significance of identified clusters by comparing the explained variance from the given clustering to what would be expected from a reference unimodal distribution. We have illustrated how the method could be applied to hierarchical clustering methods and similar methods can be developed for other clustering methods of interest. Through our simulation studies and the application to the OPPERA study we have shown that our method is a useful tool for testing the strength of identified subgroups.

In the simulation studies conducted in this paper we have found that our method compares favorably with competing methods and can even outperform other methods under certain conditions. Since our method does not require parametric assumptions it is especially useful at correctly concluding the presence or absence of clusters when the data deviates from normality. By using the information across all features our method is better able to assess the clustering structure of the data than methods which reduce the data to a univariate summary.  Unlike other non-parametric methods our method is specifically adapted for the high-dimensional setting through the use of dimension reduction techniques. Surprisingly, we found that our method outperformed SigClust in the high dimensional normally distributed clustered setting since only a portion of the features were responsible for the clusters.

One important aspect of our method is that it is agnostic to the kind of method that was used to split the data. We have illustrated how our method can be applied to test the significance of clusters produced using hierarchical clustering methods, but our method is general enough to be applied to other clustering methods as well. Also the $L_2^2$ distance used to calculate the cluster index could easily be replaced with another distance measure to accommodate different testing assumptions. 

It should be noted that our method is specifically built under the assumption that a single cluster (or unclustered data) comes from a unimodal distribution. Thus, like all clustering significance testing methods, our method is limited in its scope of when it can be useful. We have shown in this paper that our method can be useful in a variety of settings, but unfortunately it is not suited for testing clustering when the underlying data structure is naturally bimodal (such as bivariate data) or multi-modal (such as categorical data). Other testing methods should be used in those scenarios. In this paper we have illustrated that our proposed method successfully fills the missing niche of non-parametric cluster significance testing specifically adapted for high dimensional testing.  

Presently, we have focused on using our method to identify clusters which differ based on feature means. An avenue of future research would be to apply these method to identify clusters which differ based on feature variance. One possible way this could be achieved would be by decomposing the data into singular values and performing our test on the singular values instead of the original data set. Also the cluster index used as test statistic is very susceptible to outliers and future research is needed to analyze how our method can best be applied in these situations. In some situations simple preprocessing steps, such as outlier removal, can be reasonable step before testing for clustering. Future research in this topic could include using a weighted cluster index in order to down weight the effect of outliers in cluster assignment. 
\\
\bigskip
\begin{center}
	{\large\bf APPENDIX}
\end{center}

\subsection*{A1. Proof of Theorem 1} 
\citet{Ibragimov1956on}  proves that for a proper unimodal distribution function to be strong unimodal it is necessary and sufficient that the distribution function F(X) be continuous, and the function log f(x) be concave at a set of points where neither the right nor the left derivative of the function F(X) are equal to zero. Here we use the notation f(x) represents the derivative of F(x), f'(x) represents the first derivative of f(x), and f"(x) represents the second dervative.  To show that the distribution of a feature from our reference distribution $F_j(\emph{X}_j)$ is unimodal we must show that log $f_j(\emph{x}_j)$ is concave for the set of points which does not contain the mode. To achieve this we need to show that the derivative of log $f_j(\emph{X}_j)$ is negative. In the proof we appeal to the definition of univariate modality from \citet{Hartigan1985TheDip}  that a point m is a mode for the distribution function $F_j$ if the density increases on $(-\infty,m_j)$ and decreases on $(m_j,\infty)$.

textit{Proof}.
\begin{math}
\ We \ have \ that log \ F_j'(\emph{X}_j)=   \frac{dlog f_j(x_j)}{dx_j} \ = \frac{f_j'(x_j)}{f_j(x_j)}
\\ Thus \ \frac{d^2log(f_j(x_j))} {dx_j^2} =  \frac{f_j(x_j)f_j"(x_j)-f_j'(x_j)f_j'(x_j)}{f_j^2(x_j)} 
\\
f_j(x_j)>0  \ for \ (-\infty,\infty)
\\	f_j'(x_j)=0 \ at \ m_j 
\\	f_j'(x_j)>0  \ for \ (-\infty,m_j) \ and \ f_j'(x_j)<0 \ for \ (m_j,\infty)
\\	f"_j(x_j)<0 \ for \ (-\infty,m_j) \ and \ (m_j,\infty)
\end{math}

Thus we have that $\frac{d^2log(f_j(x_j))} {dx_j^2} <0$ for $(-\infty, \infty)$ except at $m_j$. Thus we have that the features are strongly unimodal. 
\subsection*{A2. Proof of Theorem 2} 
When we multiply by the Cholesky root each feature becomes the linear combination  $Y_k=g_k(X_1,...,X_p)=\Sigma_1^p(a_{ki}*X_i)$. Equivalently $X_k=g_k^{-1}(Y_1,...,Y_p)=\Sigma_{k=1}^{p}(b_{ik}*Y_i)$ We need to show that the resulting multivariate distribution $h(y_1,...,y_p)$ is multivariate unimodal.

Using the multivariate transformation of variables formula where $|J(y_1,...,y_p)|$ is the Jacobian of the transformation we have that:

\begin{math} \label{E:changeofvar}	 \underset{\boldsymbol{y}}{max} \{ h_{Y_1,...,Y_p}(y_1,...,y_p) \} =  \underset{\boldsymbol{y}}{max} \{ \ h_{X_1,...,X_p}(g_1^{-1}(y_1,...y_p),...,g_p^{-1}(y_1,...,y_p)) \times |J(y_1,...,y_p)| \}.
\end{math}

Note since each $Y_k$ is a linear combination of the $X_i$'s the Jacobean will simply be a constant value. Also,  since each $X_i$ is independent we have that 

$h_{Y_1,...,Y_p}(g_1^{-1}(y_1,...y_p),...,g_p^{-1}(y_1,...,y_p))= |J(y_1,...,y_p)|\cdot\Pi_{i=1}^{p}f_i(g_i^{-1}(y_1,...,y_p))$

To maximize $\{\Pi_{i=1}^{p}f_i(g_i^{-1}(y_1,...,y_p))\}$ we need to maximize each $f_i(g_i^{-1}(y_1,...,y_p))$ which happens at the unique mode $m_i$. Thus the solution to $\underset{\boldsymbol{y}}{max} \{ h_{Y_1,...,Y_p}(y_1,...,y_n) \}$ is the solution to the system of equations $m_i=g_i^{-1}(y_1,...,y_p)=\Sigma_{k=1}^{p}(b_{ik}*Y_k)$ for $k=1,...,p$ and $i=1,...,p$.
Since each $Y_k$ is a non-degenerate linear mapping of the $X_i's$ the solution uniquely exists. Thus we have that $h(y_1,...,y_n)$ is unimodal.
\subsection*{A3. Proof of Theorem 3} 
\citet{Huang2015statistical} show that for the choise of For the choice of $S_1$ and $S_2$ which minimizes WSS, \begin{math}
TCI_{GAUSS}=1-\frac{2}{\pi}\frac{\lambda_1}{\sum_{j=1}^p\lambda_j} 
  \end{math}.

First note that for our null distribution the density of a given feature, $y_j$ is given by:
\begin{math}
g_j(y_j)=\frac{1}{n}\sum_{i=1}^n \frac{\sqrt{1+h_j^2}}{h_j\sqrt{\lambda_j2\pi}}exp\left\{-\frac{1}{2}\left[\frac{y_j\sqrt{1+h_j^2}-\sqrt{\lambda_j}x_{ij}}{h_j\sqrt{\lambda_j}}\right]^2\right\}
\end{math}
Where $h_j$ is the minimum bandwidth such that $g(y_i)$ is unimodal and $x_{ij}$ is the observed jth feature for the ith individual. Note that each $x_{ij}$ was scaled and centered such that the sample mean was equal to 0 and the sample variance was equal to 1. 

Let $g(\boldsymbol{y})=\Pi_{j=1}^{p}g_i(y_i)$
Then the total sum of squares for $\boldsymbol{y}$ is given by:
\begin{equation}
\begin{split}
TSS=\int_{-\infty}^{\infty}\cdots\int_{-\infty}^{\infty}||\boldsymbol{y}||^2g(\boldsymbol{y})dy_1,\cdots,dy_p=
\int_{-\infty}^{\infty}\cdots\int_{-\infty}^{\infty}\sum_{j=1}^py_j^2(\Pi_{j=1}^pg_j(y_j))dy_1,\cdots,dy_p
\\
=\sum_{j=1}^p\int_{-\infty}^{\infty}y_j^2g_j(y_j)=\sum_{j=1}^p\frac{\lambda_j}{1+h_1^2}[h_1^2+2h_j\frac{1}{n}\sum_{i=1}^nx_{ij}+\frac{1}{n}\sum_{i=1}^nx_{ij}^2]=\sum_{j=1}^p\lambda_j
\end{split}
\end{equation}

Since the greatest variation is in the first feature our separating plane will be the plane which is through $\boldsymbol{\mu}=(0,\dots,0)^T$ and orthogonal to $(1,\dots,0)^T$. Let $\mu_1=(\mu_{11},...,\mu_{1p})$ By symmetry we have that $\mu_{12}=,...,=\mu_{1p}=0$. Next we need to find $\mu_{11}$.

\begin{math}
\mu_{11}=2\int_0^{\infty}y_1g_1(y_1)dy_1=\frac{2}{n}\sum_{i=1}^n\int_0^{\infty}y_1\frac{\sqrt{1+h_1^2}}{h\sqrt{\lambda_12\pi}}exp\left\{-\frac{1}{2}\left[\frac{y_1\sqrt{1+h_1^2}-\sqrt{\lambda_1}x_{i1}}{h_1\sqrt{\lambda_1}}\right]^2\right\}dy_1 
\end{math}
Where 2 is the normalization constant. Let $U=\left[\frac{\sqrt{1+h_1^2}y_1-\sqrt{\lambda_1}x_{i1}}{h_1\sqrt{\lambda_1}}\right]$. Then 
\begin{equation}
\begin{split}
\mu_{11}=\frac{2}{n}\sum_{i=1}^n\int_{-x_{i1}/h_1}^{\infty}\frac{\sqrt{\lambda_1}}{\sqrt{1+h_1^2}}[h_1U+x_{i1}]\frac{1}{\sqrt{2\pi}}exp\left\{-\frac{1}{2}U^2\right\}=
\\
\frac{2}{n}\sum_{i=1}^n\frac{\sqrt{\lambda_1}}{\sqrt{(1+h_1^2)2\pi}}\left[h_1exp\left\{-\frac{x_{i1}^2}{2h_1^2}\right\}+\int_{-x_{i1}/h_{1}}^{\infty}x_{i1}*exp\left\{-\frac{1}{2}U^2\right\}dU   \right]
\end{split}
\end{equation}
\begin{equation}
\begin{split}
\underset{n \rightarrow \infty }{\lim} \mu_{11}=
2\int_{-\infty}^{\infty}\frac{\sqrt{\lambda_1}}{\sqrt{(1+h_1^2)2\pi}}\left[h_1exp\left\{-\frac{x^2}{2h_1^2}\right\}+\int_{-x/h_{1}}^{\infty}x*exp\left\{-\frac{1}{2}U^2\right\}dU   \right]
\frac{1}{\sqrt{2\pi}}exp\left\{-\frac{1}{2}x^2\right\}dx
\\
=\frac{2\sqrt{\lambda_1}}{\sqrt{(1+h_1^2)2\pi}}\left[\int_{-\infty}^{\infty}h_1exp\left\{-{x^2}\left\{\frac{1+h_1^2}{2h_1^2}\right\}\right\}dx+
\int_{-\infty}^{\infty}\int_{-uh_1}^{\infty}\frac{x}{\sqrt{2\pi}}exp\left\{-\frac{1}{2}x^2-\frac{1}{2}U^2\right\}dx*dU\right]
\\
=\frac{2\sqrt{\lambda_1}}{\sqrt{(1+h_1^2)2\pi}}\left[\frac{h_1^2}{\sqrt{1+h_1^2}}+
\int_{-\infty}^{\infty}\frac{1}{\sqrt{2\pi}}exp\left\{-\frac{1}{2}U^2[1+h^2]\right\}dU\right]
\\=\frac{2\sqrt{\lambda_1}}{\sqrt{(1+h_1^2)2\pi}}\left[\frac{h_1^2}{\sqrt{1+h_1^2}}+
\frac{1}{\sqrt{1+h^2}}\right]
=\sqrt{\frac{2\lambda_1}{\pi}}
\end{split}
\end{equation}
Similarly $\mu_{21}=-\sqrt{\frac{2\lambda_1}{\pi}}$ and $\mu_{22}=,...,=\mu_{2p}=0$

Then we have that the within cluster sum of squares for the first cluster is given by 
\begin{equation}
\begin{split}
WSS_1=\int_{0}^{\infty}\int_{-\infty}^{\infty}\cdots\int_{-\infty}^{\infty}||\boldsymbol{y}-\boldsymbol{\mu_1}||^2g(\boldsymbol{y})dy_1,\cdots,dy_p
\\      
=\int_{0}^{\infty}(y_1-\mu_{11})^2g_1(y_1)dy_1+\sum_{j=2}^p\int_{0}^{\infty}\int_{-\infty}^{\infty}\cdots\int_{-\infty}^{\infty}y_j^2g(\boldsymbol{y})dy_1,\dots,dy_p
\end{split}
\end{equation}
\begin{equation*}
\begin{split}
\sum_{j=2}^p\int_{0}^{\infty}\int_{-\infty}^{\infty}\cdots\int_{-\infty}^{\infty}y_j^2g(\boldsymbol{y})dy_1,\dots,dy_p=\sum_{j=2}^p\int_{0}^{\infty}g_1(y_1)dy_1\left[\int_{-\infty}^{\infty}\cdots\int_{-\infty}^{\infty}y_j^2g(\boldsymbol{y})\right]dy_2,\dots,dy_p
\\
=\sum_{j=2}^p\int_{0}^{\infty}g_1(y_1)dy_1\int_{-\infty}^{\infty}y_j^2g(y_j)dy_j=\sum_{j=2}^p\frac{\lambda_j}{n}\sum_{i=1}^n\int_{0}^{\infty}\frac{\sqrt{1+h_1^2}}{h_1\sqrt{2\pi\lambda_1}}exp\{-\frac{1}{2}\left[\frac{\sqrt{1+h_1^2}y_1-\sqrt{\lambda_1}x_{i1}}{h_1\sqrt{\lambda_1}} \right]^2\}dy_1  
\end{split}
\end{equation*}
\begin{equation*}
\begin{split}
\underset{n \rightarrow \infty }{\lim} \sum_{j=2}^p\frac{\lambda_j}{n}\sum_{i=1}^n\int_{0}^{\infty}\frac{\sqrt{1+h_1^2}}{h_1\sqrt{2\pi\lambda_1}}exp\{-\frac{1}{2}\left[\frac{\sqrt{1+h_1^2}y_1-\sqrt{\lambda_1}x_{i1}}{h_1\sqrt{\lambda_1}} \right]^2\}dy_1
\\
=\sum_{j=2}^p\frac{\lambda_j}{\sqrt{2\pi}}\int_{0}^{\infty}\int_{-\infty}^\infty\frac{\sqrt{1+h_1^2}}{h_1\sqrt{2\pi\lambda_1}}exp\{-\frac{1}{2}\left[\frac{\sqrt{1+h_1^2}y_1-\sqrt{\lambda_1}x_{i1}}{h_1\sqrt{\lambda_1}} \right]^2\}exp\{-\frac{x_1^2}{2}\}dx_1dy_1
\\
=\sum_{j=2}^p\lambda_j\int_{0}^{\infty}\frac{1}{\sqrt{2\pi\lambda_1}}exp\{\frac{y_1^2}{\lambda_1}\}dy_1=\frac{1}{2}\sum_{j=2}^p\lambda_j
\end{split}
\end{equation*}
\begin{equation*}
\int_{0}^{\infty}(y_1-\mu_{11})^2g_1(y_1)dy_1=\int_{0}^{\infty}\frac{(y_1-\sqrt{\frac{2\lambda_1}{\pi}})^2}{n}\sum_{i=1}^n\int_{0}^{\infty}\frac{\sqrt{1+h_1^2}}{h_1\sqrt{2\pi\lambda_1}}exp\{-\frac{1}{2}\left[\frac{\sqrt{1+h_1^2}y_1-\sqrt{\lambda_1}x_{i1}}{h_1\sqrt{\lambda_1}} \right]^2\}dy_1
\end{equation*}
\begin{equation*}
\begin{split}
\underset{n \rightarrow \infty }{\lim}\int_{0}^{\infty}(y_1-\mu_{11})^2g_1(y_1)dy_1=
\\
=\int_{0}^{\infty}(y_1-\sqrt{\frac{2\lambda_1}{\pi}})^2\frac{\sqrt{1+h_1^2}}{h_1\sqrt{2\pi\lambda_1}}\int_{-\infty}^\infty\frac{1}{\sqrt{2\pi}}exp\{-\frac{1}{2}\left[\frac{\sqrt{1+h_1^2}y_1-\sqrt{\lambda_1}x_{1}}{h_1\sqrt{\lambda_1}} \right]^2\}exp\{-\frac{x_1^2}{2}\}dx_1dy_1
\\
=\int_{0}^{\infty}(y_1-\sqrt{\frac{2\lambda_1}{\pi}})^2\frac{1}{\sqrt{\lambda_12\pi}}exp\{-\frac{y_1^2}{2\lambda_1}\}dy_1=\frac{\lambda_1}{2}-\frac{\lambda_1}{\pi}
\end{split}
\end{equation*}
Thus we have that $WSS_1=1/2\sum_{j=1}^{p}\lambda_j-\frac{\lambda_1}{\pi}$.

We have that $WSS_1=WSS_2$ So the theoretical cluster index for our null distribution is given by
\begin{equation}
\frac{WSS_1+WSS_2}{TSS}=\frac{\sum_{j=1}^{p}\lambda_j-\frac{2\lambda_1}{\pi}}{\sum_{j=1}^p\lambda_j}=
1-\frac{2}{\pi}\frac{\lambda_1}{\sum_{j=1}^p\lambda_j}=TCI_{GAUSS}
\end{equation}
\subsection*{A4. Proof of Theorem 4} 
  Again as the number of observations, n, approaches infinity the cluster index from the data approaches the theoretical cluster index so we will show that the theoretical cluster index from the mixture distribution, $CI_{mix}$ is less than the theoretical cluster index from the null distribution, $CI_{null}$.
  First we have that that the variance for feature j in the data is given by $\lambda_j +\eta(1−\eta)a^2$. Thus we have that
  
  \begin{math}
  CI_{null}=1-\frac{2}{\pi}\frac{\lambda_1+\eta(1-\eta)a^2}{p\eta(1-\eta)a^2+\sum_{j=1}^p\lambda_j}
  \end{math}
  
  Next we determine the theoretical total sum of squares for the mixture distribution about the overall mean $\boldsymbol{\mu}=((1-\eta)a, \dots (1-\eta)a)^T$.
  
  \begin{math}
  TSS_{mix}=\int||\boldsymbol{x}-\boldsymbol{\mu}||^2 \{ \eta f(\boldsymbol{x})+(1-\eta)g(\boldsymbol{x}) \}d\boldsymbol{x}
  \\
  =\sum_{j=1}^p\int_{-\infty}^{\infty}(x_j-(1-\eta)a)^2\{\eta f(x_j)+(1-\eta)g(x_j)\}dx_j  	=p(1-\eta)\eta a^2 + \sum_{j=1}^p\lambda_j
  \end{math}
  
  The theoretical within cluster sum of squares is given by:
  
  \begin{math}
  WSS_{mix}=\int_{\boldsymbol{x}\in S_1}||\boldsymbol{x}-\boldsymbol{\mu_1}||^2 f(\boldsymbol{x})d\boldsymbol{x}
  +\int_{\boldsymbol{x}\in S_2}||\boldsymbol{x}-\boldsymbol{\mu_2}||^2 g(\boldsymbol{x}) d\boldsymbol{x}
  \end{math}
  
  Where $S_1$ and $S_2$ are partitions of $R^p$ chosen to minimize $WSS_{mix}$, $\boldsymbol{\mu_1}=
  \int_{\boldsymbol{x}\in S_1}f(\boldsymbol{x})d\boldsymbol{x}$, and $\boldsymbol{\mu_2}=
  \int_{\boldsymbol{x}\in S_2}g(\boldsymbol{x})d\boldsymbol{x}$. Replace x with $y+aI(x\in C_2)$ Where $I(x\in C_2)=1$ if $x$ is in cluster 2 and $I(x\in C_2)=0$  otherwise. $y \sim N(0,D)$. Let $h(y)$ represent the density of y.
  
  \begin{math}
  WSS_{mix}=WSS_{mix}^*=\int_{\boldsymbol{y}\in S_1^*}||\boldsymbol{y}-\boldsymbol{\mu_1^*}||^2 h(\boldsymbol{y})d\boldsymbol{y}
  +\int_{\boldsymbol{y}\in S_2^*}||\boldsymbol{y}-\boldsymbol{\mu_2^*}||^2 h(\boldsymbol{y}) d\boldsymbol{y}
  \end{math}
  
  Where here $S_1^*$ and $S_2^*$ are partitions of $R^p$ chosen to minimize $WSS_{mix}^*$, $\boldsymbol{\mu_1^*}=
  \int_{\boldsymbol{y}\in S_1^*}h(\boldsymbol{y})d\boldsymbol{y}$ and $\boldsymbol{\mu_2^*}=
  \int_{\boldsymbol{y}\in S_2^*}h(\boldsymbol{y})d\boldsymbol{y}-\boldsymbol{a}$.
  
  Since the greatest variation is in the first feature our separating plane will be the same as that for the unclustered scenario which is the plane through $(0,\dots,0)^T$ and orthogonal to $(1,\dots,0)^T$. Again we have $\mu_{12}=,...,=\mu_{1p}=\mu_{22}=,...,=\mu_{2p}=0$ and $\mu_{11}=\sqrt{\frac{2\lambda_1}{\pi}}=-\mu_{21}$. Thus we have that $WSS_{mix}=\sum_{j=1}^p\lambda_j-\frac{2\lambda_1}{\pi}$.
  Therefore
  
  \begin{math}
  TCI_{mix}=\frac{\pi\sum_{j=1}^p\lambda_j-2\lambda_1}{p(1-\eta)\eta a^2 + \sum_{j=1}^p\lambda_j}<\frac{\pi\sum_{j=1}^p\lambda_j+\pi\eta(1-\eta)a^2p-2\lambda_1-2\eta(1-\eta)a^2}{p\eta(1-\eta)a^2+\sum_{j=1}^p\lambda_j}=TCI_{null}
  \end{math}
\subsection*{A5. Proof of Theorem 5} 
  To prove this claim we employ the same strategy as \citet{liu2008statistical} by showing that 1). the cluster index from the data converges to 0 in probability as $ p \rightarrow \infty$ and 2). the cluster index under the null hypothesis is bounded away from 0 as $p \rightarrow \infty$.  We will refer to the data matrix from the reference distribution as $X^0$. Part 1). of the proof follows directly from the proof of Theorem 1 given in \citet{liu2008statistical}. For part 2). we follow a similar strategy as \citet{liu2008statistical} and use the HDLSS geometry of \citet{Hall2005Geometric}. In order to use this geometry we need to ensure that three assumptions are met: a). The fourth moments of the entry of the data vectors are uniformly bounded. b). For a constant $\sigma^2, \ lim_{p \rightarrow \infty}\frac{1}{p}\Sigma_{k=1}^pvar(X_k^0) = \sigma^2$ for features $k=1,...,p$. c). The random vector is $\rho$ mixing for functions that are dominated by quadratics. 
  
  For assumption b). We have that:
  \begin{math}
  lim_{p \rightarrow \infty}\frac{1}{p}\Sigma_{k=1}^pvar(X_k^0) =lim_{p \rightarrow \infty}\frac{1}{p}\Sigma_{k=1}^p \{\lambda_k+\eta(1-\eta)a^2\}/p=\eta(1-\eta)a^2=\sigma^2
  \end{math}
  
  For assumption a). Since n is finite we have that $max_j (\frac{1}{n}\Sigma_{i=1}^n X_{ij}^{s\ 4})<C$ for a fixed constant $C>0$. Let $K(U)$ represent the standard normal density. Since our procedure involves determining the unimodal Gaussian KDE for each feature of $X^s$ and then multiplying by the square root of the observed variance for $X$  we have that the numerical 4th moment for the Gaussian KDE of the jth feature is given by:
    \begin{math}
  \kappa_{j4}(G)=\int_{-\infty}^{\infty}x^4\frac{\sqrt{1+h_{1j}^2}}{h_{1j}\sqrt{\lambda_j+\eta(1-\eta)a^2}} \frac{1}{n}\sum_{i=1}^{n} K(\frac{x\sqrt{1+h_{1j}^2}-x_{ij}\sqrt{\lambda_j+\eta(1-\eta)a^2}}{h_{1j}\sqrt{\lambda_j+\eta(1-\eta)a^2}})dx
  \end{math} 
  
  Using the change of variable transformation $u =\frac{x\sqrt{1+h_{1j}^2}-x_{ij}\sqrt{\lambda_j+\eta(1-\eta)a^2}}{h_{1j}\sqrt{\lambda_j+\eta(1-\eta)a^2}} $ 
  and the notation that $\kappa_m(K)$ represent the mth moment for the standard Gaussian distribution. 
  
  We have: \begin{math}
  \kappa_{j4}(G)
  =\frac{(\lambda_j+\eta(1-\eta)a^2)^2}{n(1+h_{1j})^2}\sum_{i=1}^n\int_{-\infty}^{\infty}(x_{ij}^{s}+uh_{1j})^{4}K(u)du
  \\
  =\frac{(\lambda_j+\eta(1-\eta)a^2)^2}{n(1+h_{1j})^2}\sum_{i=1}^n(x_{ij}^{s\ 4}+4x_{ij}^{s\ 3}h_{1j}\kappa_1(K)+6x_{ij}^{s\ 2}(h_{1j})^2\kappa_2(K)+4x_{ij}^s(h_{1j})^3\kappa_3(K)+h_{1j}^4\kappa_4(K))
  \end{math} 
  
  Note for the Gaussian kernel we have $\kappa_1(K)=0,\kappa_2(K)=1,\kappa_3(K)=0,\kappa_4(K)=3$ also since the data has been scaled and centered we have that $\frac{1}{n}\sum_{i=1}^nx_{ij}^{s\ 2}$=1. Thus:
  
  \begin{math}
  \kappa_{j4}(G)=\frac{(\lambda_j+\eta(1-\eta)a^2)^2}{n(1+h_{1j})^2}\{\frac{1}{n}\sum_{i=1}^nx_{ij}^4+ 6h_{1j}^2+3h_{1j}^4 \}
  \\ \textup{So} \  max_i(\kappa_{i4}(G))<M^2(C+6L^2+3L^4)
  \end{math}
  
  Thus we can conclude that the fourth moments of all entries of $X^0$ are bounded uniformly. Since the entries of $X^0$ are independent we also have that assumption c). is met. Therefore the HDLSS geometry holds and have that as
  $p \rightarrow \infty,  ||X^0_j-X^0_l||^2=2p\sigma^2+O_P(1)$
  
  The proof that the CI for the null distribution $X^0$ converges away from 0 follows from the proof of 2) for theorem 1 in \citet{liu2008statistical}. The desired result then follows. \\

\bibliographystyle{ECA_jasa}
\bibliography{EHreference_all}
\end{document}